\documentclass [12pt,a4paper]{article}
\usepackage{amssymb, theorem}
\newtheorem{thm}{Theorem}

\newtheorem{lemma}{Lemma}
\theorembodyfont{\rm}
\newtheorem{pl}{Example}

\newtheorem{problem}{Problem}

\def\proof{{\it Proof: }}
\def\Cov{\mathrm{Cov}}

\def\Det{\mathrm{det}}
\def\QCov{\mathrm{qCov}}
\def\bal{\langle\!\langle}
\def\jobb{\rangle\!\rangle}

\def\qed{\nobreak\hfill $\square$}
\def\<{\langle}
\def\>{\rangle}
\def\aa{\alpha}

\def\fel{{\textstyle{1 \over 2}}}
\def\Mn{M_n(\bbbc)}
\def\iH{{\cal H}}
\def\iK{{\cal K}}
\def\iA{{\cal A}}

\def\iP{{\cal P}}

\def\bL{{\bf L}}

\def\iM{{\cal M}}
\def\iN{{\cal N}}

\def\iF{{\cal F}}

\def\Mn{M_n(\bbbc)}

\def\im{\mathrm{i}}

\def\osum{\oplus}

\def\bbbr{{\mathbb R}}
\def\bbbc{{\mathbb C}}
\def\Diag{\mbox{Diag}\,}
\def\iX{{\cal X}}

\def\Tr{\mathrm{Tr}\,}
\def\Scal{\mathrm{Scal}\,}
\def\pont{{\, \cdot \,}}
\def\eps{\varepsilon}
\def\J{{\mathbb J}}

\def\bL{{\mathbb L}}
\def\bR{{ \mathbb R}}

\def\fel{\textstyle{\frac{1}{2}}}
\def\pard{\partial}
\def\ffi{\varphi}

\begin{document}
%\rightline{\today }
\ \vskip 1cm 
\centerline{\LARGE {\bf From $f$-divergence to}}
\bigskip
\centerline{\LARGE {\bf quantum quasi-entropies and their use}}
\bigskip
\bigskip
\bigskip
\centerline{\it Dedicated to Professor Imre Csisz\'ar}
\bigskip
\bigskip
\centerline{D\'enes Petz\footnote{E-mail: petz@math.bme.hu.
Partially supported by the Hungarian Research Grant OTKA  T068258.}}
%\begin{center}
\centerline{Alfr\'ed R\'enyi Institute of Mathematics, H-1364 Budapest,
POB 127, Hungary}
\bigskip
\bigskip
\begin{abstract}
Csisz\'ar's $f$-divergence of two probability distributions was extended 
to the quantum case by the author in 1985. In the quantum setting
positive semidefinite matrices are in the place of probability 
distributions and the quantum generalization is called quasi-entropy
which is related to some other important concepts as covariance, quadratic
costs, Fisher information, Cram\'er-Rao inequality and uncertainty relation.
A conjecture about the scalar curvature of a Fisher information geometry
is explained. The described subjects are overviewed in details in the 
matrix setting, but at the very end the von Neumann algebra approach is 
sketched shortly.

\end{abstract}

Let $\iX$ be a finite space with probability measures $p$ and $q$. Their
{\bf relative entropy} or {\bf divergence}
$$
D(p||q)= \sum_{x\in \iX} p(x) \log \frac{p(x)}{q(x)}
$$
was introduced by Kullback and Leibler in 1951 \cite{K-L}. More precisely,
if $p(x)=q(x)=0$, then $\log (p(x)/q(x))=0$ and if $p(x)\ne 0$ but $q(x)=0$
for some $x \in \iX$, then $\log (p(x)/q(x))=+\infty$.

A possible generalization of the relative entropy is the $f$-divergence  
introduced by Csisz\'ar:
\begin{equation}
D_f(p||q)=\sum_{x\in \iX} q(x) f \Big(\frac{p(x)}{q(x)}\Big)
\label{relentdef}
\end{equation}
with a real function $f(x)$ defined for $x>0$ \cite {CI63,CI72}. For the
convex function $f(x)=x \log x $ the relative entropy is obtained. 

This paper first gives a rather short survey about $f$-divergence and 
we turn to the non-commutative (algebraic, or quantum) generalization. 
Roughly speaking this means that the positive $n$-tuples $p$ and $q$ are replaced
by positive semidefinite $n \times n$ matrices and the main questions 
in the study remain rather similar to the probabilistic case. The quantum 
generalization was called quasi-entropy and it is related to some other 
important concepts as covariance, quadratic costs, Fisher information, Cram\'er-Rao 
inequality and uncertainty relation. These subjects are overviewed in details in the 
matrix setting, but at the very end the von Neumann algebra approach is sketched shortly.  
When the details are not presented, the precise references are given.

\section{$f$-divergence and its use}

Let $\iF$ be the set of continuous convex functions $\bbbr^+ \to \bbbr$.
The following result explains the importance of convexity.

Let $\iA$ be a partition of $\iX$. If $p$ is a probability distribution
on $\iX$, then $p_\iA (A):= \sum_{x \in A} p(x)$ becomes a probability
distribution on $\iA$

\begin{thm}\label{T:mon1}
Let $\iA$ be a partition of  $\iX$ and $p,q$ be probability distributions 
on $\iX$. If $f\in \iF$, then
$$
D_f(p_\iA ||q_\iA)\le D_f(p||q).
$$
\end{thm}

The inequality in the theorem is the monotonicity of the $f$-divergence.
A particular case is
$$
f(1) \le D_f(p||q).
$$

\begin{thm}
Let $f, g\in \iF$ and assume that
$$
D_f(p||q) = D_g(p||q).
$$
for every distribution $p$ and $q$. Then there exists a constant $c \in \bbbr$
such that $f(x)-g(x)=c(x-1)$.
\end{thm}

Since the divergence is a kind of informational distance, we want $D_f(p||p)=0$
and require $f(1)=0$. This is nothing else but a normalization,
$$
D_{f+c}(p||q)= D_f(p||q)+c.
$$
A bit more generally, we can say that if $f(x)-g(x)$ is a linear function, then
$D_f$ and $D_g$ are essentially the same quantities.

It is interesting to remark that  $q f(p/q)$ can be considered also as a 
mean of $p$ and $q$. In that case the mean of $p$ and $p$ should be $p$,
so in the theory of means $f(1)=1$ is a different natural requirement.
 
Set $f^*(x)=x f(x^{-1})$. Then $D_f(p||q)=D_{f^*}(q||p)$. The equality 
$f^*=f$ is the symmetry condition.

\begin{pl}
Let $f(x)=|x-1|$. Then
$$
D_f(p,q)=\sum_x |p(x)-q(x)|=:V(p,q)
$$
is the {\bf variational distance} of $p$ and $q$. \qed
\end{pl} 

\begin{pl}
Let $f(x)=(1-\sqrt{x})^2/2$. Then
$$
D_f(p,q)=\sum_x (\sqrt{p(x)}-\sqrt{q(x)})^2=:H^2(p,q)
$$
is the {\bf squared Hellinger distance} of $p$ and $q$. \qed
\end{pl} 

\begin{pl}
The function
$$
f_\aa(t)={1 \over \aa(1-\aa)}\big(1-t^{\aa}\big)
$$
gives  the relative $\alpha$-entropy 
\begin{equation}\label{E:aaent}
S_\aa (p\|q)= {1 \over \aa(1-\aa)}
\Big( 1-\sum_x p(x)^{\aa}q(x)^{1-\aa}\Big). 
\end{equation}
The limit $\aa \to 0$ gives the relative entropy. \qed
\end{pl}

Several other functions appeared in the literature, we list a few of them:
\begin{equation}\label{E:fu1}
f^{(s)}(x)=\frac{1}{s(1-s)}(1+x-x^s-x^{1-s}) \qquad 0<s \ne 1 \quad \cite{CsF},
\end{equation}
\begin{equation}\label{E:fu2}
f_\beta(x)=\cases{\frac{1}{1-1/\beta}\left((1+x^\beta)^{1/\beta}-
2^{1/\beta-1}(1+x)\right) & if $0 < \beta \ne 1$,  
\cr  \phantom{MMM} \cr
(1+x) \log 2+ x \log x - (1+x)\log (x+1)& if $\beta=1$.}
\quad \cite{OV}
\end{equation}

The following result of Csisz\'ar is a characterization (or axiomatization)
of the $f$-divergence.

\begin{thm}\label{T:axiom}
Assume that a number $C(p,q)\in \bbbr$ is associated to probability
distributions on the same set $\iX$ for all finite sets $\iX$. If
\begin{itemize}
\item[(a)] $C(p,q)$ is invariant under the permutations of the basic set $\iX$.
\item[(b)] if $\iA$ is a partition of $\iX$, then
$C(p_\iA , q_\iA)\le C(p, q)$ and the equality holds if and only if 
$$
p_\iA(A) q(x)=q_\iA(A)p(x)
$$
whenever $x \in A \in \iA$,
\end{itemize}
then there exists a convex function $f:\bbbr^+ \to \bbbr$ which is continuous
at $0$ and $C(p,q)=D_f(p||q)$ for every $p$ and $q$.
\end{thm}

\section{Quantum quasi-entropy}

In the mathematical formalism of quantum mechanics, instead of $n$-tuples
of numbers one works with $n \times n$ complex matrices. They form an algebra
and this allows an algebraic approach. In this approach, a probability density 
is replaced by a positive semidefinite matrix of trace 1 which is called {\bf 
density matrix}\cite{pd2}. The eigenvalues of a density matrix give a probability
density. However, this is not the only probability density provided by
a density matrix. If we rewrite the matrix in a certain orthonormal basis,
then the diagonal element $p_1,p_2,\dots, p_n$ form a probability density.

Let $\iM$ denote the algebra of $n \times n$ matrices with complex entries.
For positive definite matrices $\rho_1, \rho_2\in \iM$, for $A \in \iM$ and 
a function $f:\bbbr^+ \to \bbbr$, the {\bf quasi-entropy} is defined as
\begin{eqnarray}\label{E:quasi}
S^A_f (\rho_1\|\rho_2)&:=& \< A \rho_2^{1/2}, f(\Delta(\rho_1/ \rho_2))
(A\rho_2^{1/2})\> \cr
&=&\Tr  \rho_2^{1/2} A^*f(\Delta(\rho_1/ \rho_2))(A\rho_2^{1/2}),
\end{eqnarray}
where $\<B,C\>:=\Tr B^*C$ is the so-called {\bf Hilbert-Schmidt inner product} and $\Delta(\rho_1/ \rho_2):\iM \to \iM$ is a linear mapping acting 
on matrices:
$$
\Delta(\rho_1/ \rho_2)A=\rho_1 A\rho_2^{-1}.
$$
This concept was introduced in \cite{PD26, PD32}, see also Chapter 7 
in \cite{OP} and it is the quantum generalization of the $f$-entropy of 
Csisz\'ar used in classical information theory (and statistics) \cite{Csi, LV}.

The monotonicity in Theorem \ref{T:mon1} is the consequence of the Jensen 
inequality. A function $f:\bbbr^+ \to \bbbr$ is called {\bf matrix concave}
if one of the following two equivalent conditions holds:
\begin{equation}
f(\lambda A +(1-\lambda) B) \ge \lambda f(A) + (1-\lambda)f(B)
\end{equation}
for every number $0 < \lambda < 1$ and for positive definite square matrices
$A$ and $B$ (of the same size). In the other condition the number $\lambda$ is (heuristically)
replaced by a matrix:
\begin{equation}
f(CAC^* +DBD^*) \ge C f(A) C^* + Df(B)D^*
\end{equation}
if $CC^*+DD^*=I$. 

A function $f:\bbbr^+ \to \bbbr$ is called {\bf matrix 
monotone} if for positive definite matrices $A \le B$ the inequality
$f(A) \le f(B)$ holds. It is interesting that a matrix monotone function is
matrix concave and a matrix concave function is matrix monotone if it is
bounded from below \cite{HP}.

Let $\alpha :\iM_0 \to \iM$ be a mapping between two matrix algebras. The dual
$\alpha^*: \iM \to \iM_0$ with respect to the Hilbert-Schmidt inner product
is positive if and only if $\alpha$ is positive. Moreover, $\alpha$ is
unital if and only if $\alpha^*$ is trace preserving. $\alpha: \iM_0 \to \iM$
is called a {\bf Schwarz mapping} if 
\begin{equation}\label{E:Sch}
\alpha(B^*B)\ge \alpha(B^*)\alpha(B)
\end{equation}
for every $B \in \iM_0$. 

The quasi-entropies are monotone and jointly convex \cite{OP, PD32}.

\begin{thm}\label{T:quasimon}
Assume that $f:\bbbr^+ \to \bbbr$ is an operator monotone function
with $f(0)\ge 0$ and $\alpha:\iM_0 \to \iM$ is a unital Schwarz mapping. 
Then
\begin{equation}\label{E:quasimonB}
S^A_f (\alpha^*(\rho_1),\alpha^*(\rho_2)) \ge
S^{\alpha(A)}_f (\rho_1,\rho_2)
\end{equation}
holds for $A \in \iM_0$ and for invertible density matrices $\rho_1$ and 
$\rho_2$ from the matrix algebra $\iM$.
\end{thm}

\proof
The proof is based on inequalities for operator monotone and
operator concave functions. First note that
$$
S^A_{f+c} (\alpha^*(\rho_1),\alpha^*(\rho_2))=
S^A_f (\alpha^*(\rho_1),\alpha^*(\rho_2))+c\,\Tr \rho_1\alpha(A^*A))
$$
and 
$$
S^{\alpha(A)}_{f+c} (\rho_1,\rho_2)=S^{\alpha(A)}_f (\rho_1,\rho_2)+
c\,\Tr \rho_1 (\alpha(A)^*\alpha(A))
$$
for a positive constant $c$. Due to the Schwarz inequality (\ref{E:Sch}),
we may assume that $f(0)=0$.

Let $\Delta:=\Delta(\rho_1/ \rho_2)$ and $\Delta_0:=\Delta(\alpha^*(\rho_1)
/ \alpha^*(\rho_2))$. The operator
\begin{equation}
VX\alpha^*(\rho_2)^{1/2}=\alpha(X)\rho_2^{1/2} \qquad (X \in \iM_0)
\end{equation}
is a contraction:
\begin{eqnarray*}
\Vert \alpha(X)\rho_2^{1/2} \Vert^2 &=& \Tr \rho_2 (\alpha(X)^* 
\alpha(X)) \cr &\le&
\Tr \rho_2 (\alpha(X^*X)= \Tr \alpha^*(\rho_2) X^*X = 
\Vert X \alpha^*(\rho_2)^{1/2}\Vert^2
\end{eqnarray*}
since the Schwarz inequality is applicable to $\alpha$. A similar simple
computation gives that
\begin{equation}
V^*\Delta V \le \Delta_0\,.
\end{equation}

Since $f$ is operator monotone, we have $f(\Delta_0)\ge f(V^*\Delta V)$. 
Recall that $f$ is operator concave, therefore $f(V^*\Delta V) \ge 
V^*f(\Delta)V$ and we conclude
\begin{equation}
f(\Delta_0) \ge V^*f(\Delta)V\,.
\end{equation}
Application to the vector $A \alpha^*(\rho_2)^{1/2}$ gives the statement. \qed

It is remarkable that for a multiplicative $\alpha$ we do not need the 
condition $f(0)\ge 0$. Moreover, $V^*\Delta V = \Delta_0$ and we do not 
need the matrix monotonicity of the function $f$. In this case the only
condition is the matrix concavity, analogously to Theorem \ref{T:mon1}.

If we apply the monotonicity (\ref{E:quasimonB}) to the embedding
$\alpha(X)=X \osum X$ of $\iM$ into $\iM \osum \iM$ and to the
densities $\rho_1=\lambda E_1\osum(1-\lambda)F_1$, $\rho_2=\lambda E_2
\osum (1-\lambda)F_2$, then we obtain the joint concavity of the 
quasi-entropy:
$$
\lambda S^A_f (E_1,E_2)+(1-\lambda) S^A_f (F_1,F_2) 
\le S^{A}_f (\lambda E_1 +(1 -\lambda)E_2)+ 
    S^A_f(\lambda F_1 +(1 -\lambda)F_2)
$$
holds. The case $f(t)=t^\alpha$ is the famous Lieb's concavity 
theorem: $\Tr A \rho^\alpha A^* \rho^{1-\alpha)}$ is concave in $\rho$ 
\cite{Lieb}. 

The concept of quasi-entropy includes some important special cases. If 
$\rho_2$ and $\rho_1$ are different and $A=I$, then we have a kind of relative 
entropy. For $f(x)= x\log x$ we have Umegaki's relative entropy 
$S(\rho_1\|\rho_2)=\Tr \rho_1 (\log \rho_1 - \log \rho_2)$. (If we want a
matrix monotone function, then we can take $f(x)=\log x$ and then we get
$S(\rho_2\|\rho_1)$.) Umegaki's relative entropy is the most important 
example, therefore the function $f$ will be chosen to be matrix convex.
This makes the probabilistic and non-commutative situation compatible
as one can see in the next argument.

Let $\rho_1$ and $\rho_2$ be density matrices in $\iM$. If in certain
basis they have diagonal $p=(p_1.p_2, \dots ,p_n)$ and $q=(q_1,q_2, 
\dots ,q_n)$, then the monotonicity theorem gives the inequality
\begin{equation}\label{E:sub}
D_f(p\|q) \le S_f(\rho_1\|\rho_2)
\end{equation}
for a matrix convex function $f$. If $\rho_1$ and $\rho_2$ commute, them we 
can take the common eigenbasis and  in (\ref{E:sub}) the equality appears. 
It is not trivial that otherwise the inequality is strict.

If $\rho_1$ and $\rho_2$ are different, then there is a choice for $p$ and $q$
such that they are different as well. Then
$$
0< D_f(p\|q) \le S_f(\rho_1\|\rho_2).
$$
Conversely, if $S_f(\rho_1\|\rho_2)=0$, then $p=q$ for every basis and this
implies $\rho_1=\rho_2$. For the relative entropy, a deeper result is known.
The {\bf Pinsker-Csisz\'ar inequality} says that
\begin{equation}\label{E:PCs}
(\|p-q\|_1)^2 \le 2 D(p\|q).
\end{equation}
This extends to the quantum case as
\begin{equation}\label{E:HO}
(\|\rho_1-\rho_2\|_1)^2 \le 2 S(\rho_1\|\rho_2),
\end{equation}
see \cite{HOT}, or \cite[Chap. 3]{pd2}.

\begin{problem}
It would be interesting to extend Theorem \ref{T:axiom} of Csisz\'ar to the 
quantum case. If we require monotonicity and specify the condition for equality, 
then a function $f$ is provided by Theorem \ref{T:axiom}, but for non-commuting 
densities the conclusion is not clear.
\end{problem}

\begin{pl}
Let
$$
f_\alpha(x)={1 \over \aa(1-\aa)}\big(1-x^{\aa}\big),
$$ 
is matrix monotone decreasing for $\alpha \in (-1,1)$. (For $\aa=0$, the 
limit is taken and it is $- \log x$.) Then the {\bf relative entropies of degree}
$\alpha$ are produced: 
$$
S_\alpha(\rho_2\|\rho_1):={1 \over \aa(1-\aa)}
\Tr (I-\rho_1^{\aa}\rho_2^{-\aa})\rho_2.
$$
These quantities are essential in the quantum case. \qed
\end{pl}

If $\rho_2=\rho_1=\rho$ and $A, B \in \iM$ are arbitrary, then one can 
approach  to the {\bf generalized covariance} \cite{PD22}.
\begin{equation}\label{E:qC}
\QCov^f_{\rho}(A,B):=\< A \rho ^{1/2}, f(\Delta(\rho / \rho ))(B\rho ^{1/2})\>
-(\Tr \rho A^*)(\Tr \rho B).
\end{equation}
is a generalized covariance. If $\rho, A$ and $B$ commute, then this 
becomes $f(1) \Tr \rho A^*B-(\Tr \rho A^*)(\Tr \rho B)$. This shows that
the normalization $f(1)=1$ is natural. The generalized covariance 
$\QCov^f_{\rho}(A,B)$ is a sesquilinear form and it is determined by 
$\QCov^f_{\rho}(A,A)$ when $\{ A\in \iM: \Tr \rho A=0\}$. Formally, 
this is a quasi-entropy and Theorem \ref{T:quasimon} applies if 
$f$ is matrix monotone. If we require the symmetry condition 
$\QCov^f_{\rho}(A,A)=\QCov^f_{\rho}(A^*,A^*)$, then $f$ should have
the symmetry $xf(x^{-1})=f(x)$.

Assume that $\Tr \rho A=\Tr \rho B=0$ and $\rho=\Diag(\lambda_1, \lambda_2,\dots,
\lambda_n)$. Then
\begin{equation}\label{E:qC2}
\QCov^f_{\rho}(A,B)=\sum_{ij} \lambda_i f (\lambda_j/\lambda_i)
A^*_{ij}B_{ij}.
\end{equation}

A matrix monotone function $f:\bbbr^+ \to \bbbr^+$ will be called 
{\bf standard} if $xf(x^{-1})=f(x)$ and $f(1)=1$. A standard function $f$ 
admits a canonical representation
\begin{equation}\label{E:canonicalf}
f(t)=\frac{1+t}{2}\exp
\int_0^1(1-t^2)\frac{\lambda-1}{(\lambda+t)(1+\lambda t)(\lambda+1)
}h(\lambda)\,d\lambda,
\end{equation}
where $ h:[0,1]\to[0,1] $ is a measurable function \cite{H1}.  

The usual {\bf symmetrized covariance} corresponds to the function 
$f(t)=(t+1)/2$:
$$
\Cov_\rho (A,B):=
\frac{1}{2}\Tr (\rho (A^*B+BA^*))- (\Tr \rho A^*)(\Tr \rho B).
$$

The interpretation of the covariances is not at all clear. In the next
section they will be called {\bf quadratic cost functions}. It turns out that
there is a one-to-one correspondence between quadratic cost functions
and Fisher informations.

\section{Fisher information}
\subsection{The Cram\'er-Rao inequality}

\indent
The Cram\'er-Rao inequality belongs to the basics of estimation theory
in mathematical statistics. Its quantum analog was discovered
immediately after the foundation of mathematical quantum estimation
theory in the 1960's, see the book \cite{He} of Helstrom, or the
book \cite{Ho} of Holevo for a rigorous summary of the subject. Although
both the classical Cram\'er-Rao inequality and its quantum analog 
are as trivial as the Schwarz inequality, the subject takes a lot of attention
because it is located on the highly exciting boundary of statistics,
information and quantum theory.

As a starting point we give a very general form of the quantum
Cram\'er-Rao inequality in the simple setting of finite dimensional
quantum mechanics. For $\theta\in (-\eps, \eps)\subset \bbbr$ a statistical
operator $\rho(\theta)$ is given and the aim is to estimate the value
of the parameter $\theta$ close to $0$. Formally $\rho(\theta)$ is an $n
\times n$ positive semidefinite matrix of trace 1 which describes a
mixed state of a quantum mechanical system and we assume that $\rho(\theta)$
is smooth (in $\theta$). Assume that an estimation
is performed by the measurement of a self-adjoint matrix $A$ playing the
role of an observable. $A$ is called {\bf locally unbiased estimator} if
\begin{equation}\label{E:lue}
\frac{\partial}{\partial \theta}\Tr \rho(\theta) A\Big|_{\theta=0}=1\,.
\end{equation}
This condition holds if $A$ is an {unbiased estimator} for $\theta$,
that is
\begin{equation}
\Tr \rho(\theta) A =\theta \qquad (\theta \in (-\eps,\eps)).
\end{equation}
To require this equality for all values of the parameter is a serious
restriction on the observable $A$ and we prefer to use the weaker
condition (\ref{E:lue}).

Let $\ffi_0[K,L]$ be an inner product (or quadratic cost function) on 
the linear space of self-adjoint matrices. When $\rho(\theta)$ is smooth in 
$\theta$, as already was assumed above, then
\begin{equation}\label{E:func}
\frac{\partial}{\partial \theta}\Tr \rho(\theta) B\Big|_{\theta=0}=\ffi_0[B,L]
\end{equation}
with some $L=L^*$. From (\ref{E:lue}) and (\ref{E:func}), we have
$\ffi_0[A,L]=1$ and the Schwarz inequality yields
\begin{equation}\label{E:CR}
\ffi_0[A,A] \ge \frac{1}{\ffi_0[L,L]}\,.
\end{equation}
This is the celebrated {\bf inequality of Cram\'er-Rao type} for the locally
unbiased estimator.

The right-hand-side of (\ref{E:CR}) is independent of the estimator
and provides a lower bound for the quadratic cost. The
denominator $\ffi_0[L,L]$ appears to be in the role of Fisher information
here. We call it {quantum Fisher information} with respect to the
cost function $\ffi_0[\pont,\pont]$. This quantity depends on the
tangent of the curve $\rho(\theta)$. If the densities $\rho(\theta)$ and
the estimator $A$ commute, then
\begin{equation}\label{E:comm}
L=\rho_0^{-1}\frac{d \rho(\theta)}{d \theta}\quad \mbox{and}\quad
\ffi_0[L,L]=\Tr \rho_0^{-1}\left(\frac{d \rho(\theta)}{d \theta}\right)^2=
\Tr \rho_0\left(\rho_0^{-1}\frac{d \rho(\theta)}{d \theta}\right)^2.
\end{equation}

We want to conclude from the above argument that whatever Fisher
information and generalized variance are in the quantum mechanical
setting, they are very strongly related. In an earlier work
\cite{PD2, PD3} we used a monotonicity condition to make a
limitation on the class of Riemannian metrics on the state space
of a quantum system. The monotone metrics are called Fisher
information quantities in this paper.

Since the sufficient and necessary condition for the equality in the
Schwarz inequality is well-known, we are able to analyze the case of
equality in (\ref{E:CR}). The condition for equality is
$$
A=\lambda L
$$
for some constant $\lambda \in \bbbr$.
Therefore the necessary and sufficient condition for equality in
(\ref{E:CR}) is
\begin{equation}\label{E:EQ}
\dot{\rho}_0:=\frac{\partial}{\partial \theta} \rho(\theta)
\Big|_{\theta=0}=\lambda^{-1} \J_0(A)\,.
\end{equation}
Therefore there exists a unique locally unbiased estimator $A=\lambda\J_0^{-1}
(\dot{\rho}_0)$, where the number $\lambda$ is chosen such a way that the
condition (\ref{E:lue}) should be satisfied.

\begin{pl}\label{Ex:1}
Let
$$
\rho(\theta):= \rho+\theta B,
$$
where $\rho$ is a positive definite density and $B$ is a self-adjoint
traceless operator. $A$ is locally unbiased when $\Tr AB=1$. In particular,
$$
A=\frac{B}{\Tr B^2}
$$
is a locally unbiased estimator and in the Cram\'er-Rao inequality $(\ref{E:CR})$
the equality holds when $\ffi_0[X,Y]=\Tr XY$, that is, $\J_0$
is the identity. 

If $\Tr \rho B=0$ holds in addition, then the estimator
is unbiased. \qed
\end{pl}

\subsection{Coarse-graining and monotonicity}

In the simple setting in which the state is described by a density
matrix, a {\bf coarse-graining} is an affine mapping sending density matrices
into density matrices. Such a mapping extends to all matrices and
provides a positivity and trace preserving linear transformation.
A common example of coarse-graining sends the density matrix $\rho_{12}$
of a composite system $1+2$ into the (reduced) density matrix $\rho_1$ of
component 1. There are several reasons to assume completely positivity
about a coarse graining and we do so.

Assume that $\rho(\theta)$ is a smooth curve of density matrices with
tangent $A:=\dot{\rho}$ at $\rho$. The quantum Fisher information
$F_\rho(A)$ is an information quantity associated with the pair $(\rho, A)$,
it appeared in the Cram\'er-Rao inequality above and the classical Fisher
information gives a bound for the variance of a locally
unbiased estimator. Let now $\beta$ be a coarse-graining. Then $\beta
(\rho(\theta))$ is another curve in the state space. Due to the linearity of
$\beta$, the tangent at $\beta(\rho_0)$ is $\beta(A)$. As it is usual in
statistics, information cannot be gained by coarse graining, therefore
we expect that the Fisher information at the density matrix $\rho_0$ in
the direction $A$ must be larger than the Fisher information at
$\beta(\rho_0)$ in the direction $\beta(A)$. This is the {monotonicity
property of the Fisher information} under coarse-graining:
\begin{equation}\label{E:FM}
F_\rho(A) \ge F_{\beta(\rho)}(\beta(A))
\end{equation}
Although we do not want to have a concrete formula for the quantum
Fisher information, we require that this monotonicity condition
must hold. Another requirement is that $F_\rho(A)$ should be
quadratic in $A$, in other words there exists a non-degenerate real
bilinear form $\gamma_\rho(A,B)$ on the self-adjoint matrices such
that
\begin{equation}\label{E:BFM}
F_\rho(A)=\gamma_\rho(A,A).
\end{equation}
The requirements (\ref{E:FM}) and (\ref{E:BFM}) are strong enough
to obtain a reasonable but still wide class of possible quantum Fisher
informations.

We may assume that
\begin{equation}\label{E:Jdef}
\gamma_\rho(A,B)=\Tr A\J_\rho^{-1}(B^*).
\end{equation}
for an operator $\J_\rho$ acting on matrices. (This formula
expresses the inner product $\gamma_D$ by means of the
Hilbert-Schmidt inner product and the positive linear
operator $\J_\rho$.) In terms of the operator $\J_\rho$ the
monotonicity condition reads as
\begin{equation}\label{E:Fmon}
\beta^* \J_{\beta(\rho)}^{-1}\beta  \le \J_\rho^{-1}
\end{equation}
for every coarse graining $\beta$. ($\beta^*$ stand for the adjoint
of $\beta$ with respect to the Hilbert-Schmidt product. Recall that
$\beta$ is completely positive and trace preserving if and only if
$\beta^*$ is completely positive and unital.) On the other hand the latter
condition is equivalent to
\begin{equation}\label{E:fimon1}
\beta \J_{\rho}\beta^*  \le \J_{\beta(\rho)}\, .
\end{equation}

We proved the following theorem in \cite{PD2}.

\begin{thm}\label{T:mon}
If for every invertible density matrix $\rho\in \Mn$ a positive definite sesquilinear form
$\gamma_\rho:\Mn \times \Mn \to \bbbc $ is given such that 
\begin{itemize}
\item [(1)] the monotonicity
$$\gamma_\rho(A, A) \ge \gamma_{\beta(\rho)}(\beta(A), \beta(A))
$$
holds for all completely positive coarse grainings $\beta: \Mn \to M_m(\bbbc)$,
\item [(2)] $\gamma_\rho(A,A)$ is continuous in $\rho$ for every fixed $A$,
\item [(3)] $\gamma_\rho(A,A)=\gamma_\rho(A^*,A^*)$,
\item [(4)] $\gamma_\rho(A,A)=\Tr \rho^{-1}A^2$ if $A$ is self-adjoint and 
$A\rho=\rho A$,
\end{itemize}
then there exists a unique standard operator monotone function 
$f: \bbbr^+\to \bbbr$ such that 
$$
\gamma_\rho^f(A,A)=\Tr A\J_\rho^{-1}(A)\qquad\mbox{and}\qquad
\J_\rho=\bR_\rho^{1/2}f(\bL_\rho\bR_\rho^{-1})\bR_\rho^{1/2}\,,
$$
where the linear transformations $\bL_\rho$ and  $\bR_\rho$ acting on matrices
are the left and right multiplications, that is
$$
\bL_\rho(X)=\rho X \qquad\mbox{and}\qquad \bR_\rho(X)=X\rho\,.
$$
\end{thm}

The above $\gamma_\rho(A,A)$ is formally a quasi-entropy, 
$S_{1/f}^{A\rho^{-1}}(\rho,\rho)$, however this form is not suitable
to show the monotonicity. Assume that $\rho=\Diag(\lambda_1, \lambda_2,\dots,
\lambda_n)$. Then
\begin{equation}\label{E:qF2}
\gamma^f_{\rho}(A,A)=\sum_{ij} \frac{1}{\lambda_i f (\lambda_j/\lambda_i)}
|A_{ij}|^2.
\end{equation}

It is clear from this formula that the Fisher information is affine
in the function $1/f$. Therefore, {\bf Hansen's canonical representation} 
of the reciprocal of a standard operator monotone function can be used 
\cite{H}.

\begin{thm}\label{T:hansen}
If $f:\bbbr^+ \to \bbbr^+$ be a standard operator monotone function, then
$$
\frac{1}{f(t)}=
\int_0^1 \frac{1+\lambda}{2}
\left(\frac{1}{t+\lambda}+\frac{1}{1+t\lambda}\right)d\mu(\lambda),
$$
where $ \mu $ is a probability measure on $[0,1]$.
\end{thm}

The theorem implies that the set $\{1/f: f \mbox{\ is\ standard\ operator
\ monotone}\}$ is convex and gives the extremal points
\begin{equation}\label{E:efl}
g_\lambda(x):=\frac{1+\lambda}{2}
\left(\frac{1}{t+\lambda}+\frac{1}{1+t\lambda}\right) 
\qquad (0 \le \lambda \le 1).
\end{equation}
One can compute directly that
$$
\frac{\partial}{\partial\lambda}g_\lambda(x)=
-\frac{(1-\lambda^{2})(x+1)(x-1)^{2}}{2(x+\lambda)^{2}(1+x\lambda)^{2}}.
$$
Hence $g_\lambda$ is decreasing in the parameter $\lambda$. For $\lambda=0$
we have the largest function $g_0(t)=(t+1)/(2t)$ and for $\lambda=1$ the 
smallest is $g_1(t)=2/(t+1)$. (Note that this was also obtained in the 
setting of positive operator means \cite{Ando}, harmonic and arithmetic means.)

Via the operator $\J_\rho$, each monotone Fisher information determines
a quantity
\begin{equation}\label{E:fi}
\varphi_\rho[A,A]:= \Tr A\J_\rho(A)
\end{equation}
which is a {quadratic cost functional}. According to
(\ref{E:fimon1}) (or Theorem \ref{T:quasimon}) this possesses the 
monotonicity property
\begin{equation}\label{E:fimon2}
\varphi_\rho[\beta^*(A),\beta^*(A)] \le \varphi_{\beta(\rho)}[A, A]\, .
\end{equation}
Since (\ref{E:Fmon}) and (\ref{E:fimon1}) are equivalent we observe a
one-to-one correspondence between monotone Fisher informations and
monotone quadratic cost functions. 

\begin{thm}\label{T:mon2}
If for every invertible density matrix $\rho\in \Mn$ a positive definite sesquilinear form
$\varphi_\rho:\Mn \times \Mn \to \bbbc $ is given such that 
\begin{itemize}
\item [(1)] the monotonicity (\ref{E:fimon2})
holds for all completely positive coarse grainings $\beta: \Mn \to M_m(\bbbc)$,
\item [(2)] $\varphi_\rho[A,A]$ is continuous in $\rho$ for every fixed $A$,
\item [(3)] $\varphi_\rho[A,A]=\varphi_\rho[A^*,A^*]$,
\item [(4)] $\varphi_\rho[A,A]=\Tr \rho A^2$ if $A$ is self-adjoint and 
$A\rho=\rho A$,
\end{itemize}
then there exists a unique standard operator monotone function 
$f: \bbbr^+\to \bbbr$ such that 
$$
\varphi_\rho^f[A,A]=\Tr A\J_\rho (A)
$$
with the operator $\J_\rho$ defined in Theorem \ref{T:mon}.
\end{thm}

Any such cost function has the property $\varphi_{\rho}[A, B]=\Tr \rho A^*B$ when
$\rho$ commutes with $A$ and $B$. The examples below show that it is not so 
generally.

\begin{pl}\label{Pl:symlog}
Among the standard operator monotone functions, $f_a(t)=(1+t)/2$ is maximal.
This leads to the fact that among all monotone quantum Fisher informations 
there is a smallest one which corresponds to the function $f_a(t)$. In this 
case
\begin{equation}\label{E:minF}
F_\rho^{\min}(A)=\Tr A L=\Tr \rho L^2,\qquad \mbox{where}\qquad
\rho L+L\rho=2A.
\end{equation}
For the purpose of a quantum Cram\'er-Rao inequality the minimal quantity
seems to be the best, since the inverse gives the largest lower bound.
In fact, the matrix $L$ has been used for a long time under the name
of {\bf symmetric logarithmic derivative}, see \cite{Ho} and \cite{He}.
In this example the quadratic cost function is
\begin{equation}\label{E:minFv}
\ffi_\rho[A,B] =\fel \Tr \rho(AB+BA)
\end{equation}
and we have
\begin{equation}\label{E:minFj}
\J_\rho(B)=\fel(\rho B+B\rho)\qquad \mbox{and} \qquad
\J_\rho^{-1}(A)=\int_0^\infty e^{-t\rho/2}Ae^{-t\rho/2}\,dt
\end{equation}
for the operator $\J$ of the previous section. 

To see the second formula of (\ref{E:minFj}), set $A(t):=e^{-t\rho/2}Ae^{-t\rho/2}$. 
Then
$$
\frac{d}{dt} A(t)=-\fel( \rho A(t)-A(t) \rho)
$$
and
$$
\int_0^\infty \fel(\rho A(t)+A(t)\rho )\,dt=
\big[-A(t)\big]_0^\infty=A.
$$
Hence
$$
\J_\rho\Big(\int_0^\infty A(t)\,dt\Big)=A.
$$

Let $T=T^*$ and $\rho_0$ be a density matrix. Then $D(\theta):= \exp (\theta T/2)\rho_0
\exp (\theta T/2)$ satisfies the differential equation
\begin{equation}\label{E:mindif}
\frac{\pard}{\pard \theta} D(\theta)=\J_{D(\theta)} T
\end{equation}
and
\begin{equation}\label{E:minexp}
\rho(\theta)=\frac{D(\theta)}{\Tr D(\theta)}
\end{equation}
is a kind of {\bf exponential family}. 

If $\Tr \rho_0 T=0$ and $\Tr \rho_0 T^2=1$, then
$$
\frac{\pard}{\pard \theta} \Tr \rho(\theta) T\Big|_{\theta=0}=1
$$
and $T$ is a locally unbiased estimator (of the parameter $\theta$ at $\theta=0$).
Since 
$$
\frac{\pard}{\pard \theta} \rho(\theta) \Big|_{\theta=0}=\J_0(T),
$$
we have equality in the Cram\'er-Rao inequality, see (\ref{E:EQ}). \qed
\end{pl}

\begin{pl}
The function
\begin{equation} \label{E:efek}
f_\beta(t)=\beta(1-\beta)\frac{(x-1)^2}{(x^\beta-1)(x^{1-\beta}-1)}
\end{equation}
is operator monotone if $0 < |\beta| < 1$.

When $A=\im [\rho,B]$ is orthogonal to the commutator of the foot-point
$\rho$ in the tangent space, we have
\begin{equation} \label{E:WYD}
F^\beta_\rho(A)=\frac{1}{2 \beta(1-\beta)}\Tr\big([\rho^\beta,B][\rho^{1-\beta}, 
B]\big).
\end{equation}
Apart from a constant factor this expression is the skew information
proposed by Wigner and Yanase some time ago (\cite{WYD}). In
the limiting cases $\beta \to 0$ or $1$ we have
$$
f_0(x)=\frac{x-1}{\log x}
$$ 
and the corresponding Fisher information
\begin{equation} \label{E:KM}
\gamma_\rho(A,B):=  \int_0^\infty \Tr A (\rho+t)^{-1}B(\rho+t)^{-1}\,dt
\end{equation}
is named after Kubo, Mori, Bogoliubov etc. The Kubo-Mori inner product
plays a role in quantum statistical mechanics (see \cite{FS}, for
example). In this case
\begin{equation} \label{E:KMJ}
\J^{-1}(B)=  \int_0^\infty (\rho+t)^{-1}B(\rho+t)^{-1}\,dt\quad\hbox{and}\quad
\J(A)=  \int_0^1 \rho^{t}A \rho^{1-t}\,dt\,.
\end{equation}
Therefore the corresponding quadratic cost functional is
\begin{equation} \label{E:KMvar}
\varphi_\rho[A,B]=  \int_0^1 \Tr A \rho^{t}B\rho^{1-t}\,dt\,.
\end{equation}

Let
\begin{equation} \label{E:KMexp}
\rho(\theta):= \frac{\exp (H+\theta T)}{\Tr\exp (H+\theta T)},
\end{equation}
where $\rho=e^H$. Assume that $\Tr e^H T=0$. The Frechet
derivative of $e^H$ is $\int_0^1 \Tr e^{tH} T e^{(1-t)H}\,dt$. Hence
$A$ is locally unbiased if
$$
\int_0^1 \Tr \rho^t T \rho^{1-t}A\,dt=1.
$$
This holds if
$$
A=\frac{T}{\int_0^1 \Tr \rho^t T \rho^{1-t}T\,dt}.
$$
In the Cram\'er-Rao inequality $(\ref{E:CR})$ the equality holds when 
$\J_0(K)=\int_0^1 D^t K D^{1-t}\,dt$.

Note that (\ref{E:KMexp}) is again an {\bf exponential family}, the differential 
equation for 
$$
D(\theta)=\exp (H+\theta T)
$$
has the form (\ref{E:mindif}) with 
$$
\J_{D(\theta)}(K)=\int_0^1 D(\theta)^t K D(\theta)^{1-t}\,dt.
$$
\phantom{MMMMMMMM}\qed 
\end{pl}

\begin{problem}
It would be interesting to find more exponential families. This means solution of 
the differential equation
$$
\frac{\pard}{\pard \theta} D(\theta)=\J_{D(\theta)} T, \qquad D(0)=\rho_0.
$$
If the self-adjoint $T$ and the positive $\rho$ commute, then the solution
is $D(\theta)=\exp(\theta T)\rho_0$. A concrete example is
$$
\frac{\pard}{\pard \theta} D(\theta)=D(\theta)^{1/2} T D(\theta)^{1/2}.
$$
\end{problem}

\subsection{Manifolds of density matrices}

Let $\mathcal{M} :=\{\rho(\theta): \theta \in G\}$ be a smooth 
$m$-dimensional manifold of invertible density matrices. When a quadratic
cost function $\ffi_0$ is fixed, the corresponding Fisher information
is a Riemannian metric on the manifold. This gives a possibility for
geometric interpretation of statistical statements \cite{Am, AN}.

Fisher information appears not only as a Riemannian metric but as an
information matrix as well.  The {\bf quantum score operators} (or 
logarithmic derivatives) are
defined as
\begin{equation}
L_i(\theta):= \J_{\rho(\theta)}^{-1}\big(\pard_{\theta_i} \rho(\theta)\big)
\qquad (1 \le i \le m)
\end{equation}
and
\begin{equation}
I_{ij}^Q(\theta):=\Tr L_i(\theta)\J_{\rho(\theta)}\big(L_j(\theta)\big)
\qquad (1 \le i,j \le m)
\end{equation}
is the {\bf quantum Fisher information matrix}.

The next result is the monotonicity of Fisher information matrix.

\begin{thm}{\bf \cite{PD22}}\label{T:mon3}
Let $\beta$ be a coarse-graining sending density matrices on the Hilbert
space $\mathcal{H}_1$ into those acting on the Hilbert space $\mathcal{H}_2$
and
let $\mathcal{M}:=\{\rho(\theta): \theta \in G\}$ be a smooth $m$-dimensional
manifold of invertible density matrices on $\mathcal{H}_1$. For the Fisher
information matrix $I^{1Q}(\theta)$ of $\mathcal{M}$ and for Fisher information
matrix $I^{2Q}(\theta)$ of $\beta(\mathcal{M}):=\{\beta(\rho(\theta)): \theta
\in G\}$
we have the monotonicity relation
\begin{equation}\label{E:monr}
I^{2Q}(\theta) \le I^{1Q}(\theta).
\end{equation}
\end{thm}

Assume that $F_j$ are positive operators acting on a Hilbert space
$\mathcal{H}_1$ on which the family $\mathcal{M}:=\{\rho(\theta): \theta \in
G\}$
is given. When $\sum_{j=1}^n F_j =I$, these operators determine a measurement.
For any $\rho(\theta)$ the formula
$$
\beta (\rho(\theta)):= \Diag (\Tr \rho(\theta)F_1, \dots, \Tr \rho(\theta)F_n)
$$
gives a diagonal density matrix. Since this family is commutative, all
quantum Fisher informations coincide with the classical (\ref{E:comm})
and the classical Fisher information stand on the left-hand-side of
(\ref{E:monr}). The right-hand-side can be arbitrary quantum quantity
but it is minimal if it based on the symmetric logarithmic derivative,
see Example \ref{Pl:symlog}. This particular case of the Theorem is 
in the paper \cite{BC}.

Assume that a manifold $\iM:=\{\rho(\theta): \theta \in G\}$ of density 
matrices is given together a statistically relevant Riemannian metric 
$\gamma$. Given two points on the manifold their geodesic distance is 
interpreted as the statistical distinguish-ability of the two density 
matrices in some statistical procedure.

Let $\rho_0 \in \iM$ be a point on our statistical manifold. The geodesic ball
$$
B_\eps (\rho_0):=\{\rho \in \iM: d(\rho_0,\rho) < \eps\}
$$
contains all density matrices which can be distinguished by an effort smaller
than $\eps$ from the fixed density $\rho_0$. The size of the inference region
$B_\eps (\rho_0)$  measures the statistical uncertainty at the density 
$\rho_0$. Following {\bf Jeffrey's rule} the size is the volume measure 
determined by the statistical (or information) metric. More precisely, it is 
better to consider the asymptotics of the volume of $B_\eps (\rho_0)$ as 
$\eps \to 0$. It is known in differential geometry that
\begin{equation}
Vol \big(B_\eps (\rho_0)\big)=C_m \eps^m -\frac{C_m}{6(m+2)}
\Scal (\rho_0)\eps^{m+2}+ o(\eps^{m+2}),
\end{equation}
where $m$ is the dimension of our manifold, $C_m$ is a constant (equals to the
volume of the unit ball in the Euclidean $m$-space) and $Scal$ means the scalar
curvature, see \cite[3.98 Theorem]{GHL}. In this way, the scalar curvature 
of a statistically relevant Riemannian metric might be interpreted as the 
{\bf average statistical uncertainty} of the density matrix (in the given 
statistical manifold). This interpretation becomes particularly interesting
for the full state space endowed by the Kubo-Mori inner product as a 
statistically relevant Riemannian metric.

The Kubo-Mori (or Bogoliubov) inner product is given by
\begin{equation}
\gamma_\rho(A,B)=\Tr (\partial_A \rho)( \partial_B \log \rho),
\end{equation}
or (\ref{E:KM}) in the affine parametrization. On the basis of numerical 
evidences it was {\bf conjectured} in \cite{PD1} that the scalar curvature 
which is a statistical uncertainty is monotone in the following sense. For 
any coarse graining $\alpha$ the scalar curvature at a density $\rho$ is 
smaller than at $\alpha(\rho)$. The average statistical uncertainty is 
increasing under coarse graining. Up to now this conjecture has not been 
proven mathematically. Another form of the conjecture is the statement that 
along a curve of Gibbs states
$$
\frac{ e^{-\beta H}}{\Tr e^{-\beta H}}
$$
the scalar curvature changes monotonly with the inverse temperature 
$\beta \ge 0$, that is, {\bf the scalar curvature is monotone decreasing 
function of $\beta$}. (Some partial results are in \cite{And1}.)

Let $\iM$ be the manifold of all invertible $n \times n$ density matrices.
If we use the affine parametrization, then the tangent space $T_\rho$
consists of the traceless self-adjoint matrices and has ab orthogonal
decomposition
\begin{equation}\label{E:dec}
T_\rho= \{\im [\rho, B]: B \in M_n^{sa}\} \osum \{A=A^*: \Tr A=0, \quad
A\rho= \rho A\}.
\end{equation}
We denote the two subspaces by $T_\rho^q$ and $T_\rho^c$, respectively. If
$A_2 \in T_\rho^c$, then 
$$
F(\Delta(\rho/ \rho))(A_2\rho^{\pm 1/2})=A_2\rho^{\pm 1/2}
$$
implies
$$
\QCov^f_{\rho}(A_1,A_2)=\Tr \rho A_1^*A_2 -(\Tr \rho A_1^*)(\Tr \rho A_2),
\qquad
\gamma_\rho^f (A_1,A_2)=\Tr \rho^{-1}A_1^*A_2
$$
independently of the function $f$. Moreover, if $A_1 \in T_\rho^q$, then
$$
\gamma_\rho^f (A_1,A_2)=\QCov^f_{\rho}(A_1,A_2) =0 \,.
$$
Therefore, the decomposition (\ref{E:dec}) is orthogonal with respect to
any Fisher information and any quadratic cost functional. Moreover, the
effect of the function $f$ and the really quantum situation
are provided by the components from $T_\rho^q$.

\subsection{Skew information}

Let $f$ be a standard function and $X=X^*\in M_n$. The quantity
$$
I_\rho ^f(X):= \frac{f(0)}{2}\gamma_\rho ^f(\im [\rho ,X], \im [\rho ,X] )
$$
was called {\bf skew information} in \cite{H} in this general setting.
The skew information is nothing else but the Fisher information restricted
to $T_\rho ^q$, but it is parametrized by the commutator. 

If $\rho =\Diag (\lambda_1, \dots, \lambda_n)$ is diagonal, then
$$
\gamma_\rho ^f(\im [\rho ,X], \im [\rho ,X] )=\sum_{ij} 
\frac{(\lambda_i -\lambda_j)^2}{\lambda_j f(\lambda_i/\lambda_j)}|X_{ij}|^2.
$$
This implies that the identity
\begin{equation}\label{E:tilde2}
f(0)\gamma_\rho ^f(\im [\rho ,X], \im [\rho ,X] )=
2\Cov_\rho (X,X)-2\QCov^{\tilde f}_\rho  (X, X )
\end{equation}
holds if $\Tr \rho X=0$ and
\begin{equation}\label{E:tilde1}
\tilde{f}(x):=\frac{1}{2}\left((x+1)-(x-1)^2 \frac{f(0)}{f(x)}\right).
\end{equation}

The following result was obtained in \cite{3}. 

\begin{thm}\label{T:tilda}
If $f:\bbbr^+ \to \bbbr$ is a standard function, then $\tilde{f}$ is 
standard as well.
\end{thm}

The original proof is not easy, even matrix convexity of functions of 
two variables is used. Here we sketch a rather elementary proof based 
on the fact that $1/f  \mapsto \tilde{f}$ is linear and on the canonical 
decomposition in Theorem \ref{T:hansen}.

\begin{lemma}
Let $0 \le \lambda \le 0$ and $f_\lambda:\bbbr^+ \to \bbbr$ be a function 
such that
$$
\frac{1}{f_\lambda(x)}:=
\frac{1+\lambda}{2}\left(\frac{1}{x+\lambda}+\frac{1}{1+x\lambda}\right)
=g_\lambda(x).
$$
Then the function $\widetilde{f}:\bbbr^+ \to \bbbr$ defined in (\ref{E:tilde1})
is an operator monotone standard function.
\end{lemma}
 
The proof of the lemma is elementary. From the lemma and Theorem 
\ref{T:hansen}, Theorem \ref{T:tilda} follows straightforwardly \cite{128skew}.

The skew information is the Hessian of a quasi-entropy:

\begin{thm}
Assume that $X=X^* \in M_n$ and $\Tr \rho  X=0$. If $f$ is a standard function
such that $f(0)\ne 0$, then
$$
\frac{\pard^2}{\pard t \pard s}  S_F (\rho +t\im [\rho ,X],\rho +s\im [\rho ,X])\Big|_{t=s=0}
=f(0)\gamma_\rho ^f(\im [\rho ,X], \im [\rho ,X] )
$$
for the standard function $F=\tilde f$.
\end{thm}

The proof is based on the formula
$$
{d\over dt} h(\rho +t\im [\rho,X])\Big|_{t=0} =\im [h(\rho),X]\,,
$$
see \cite{128skew}.

\begin{pl}
We compute the Hessian of the relative entropy of degree $\alpha$ in an 
exponential parametrization:
$$
\frac{\partial^2}{\partial t \partial s}S_\aa (e^{H+tA}|| e^{H+sB})
\Big|_{t=s=0}=\int_0^1 \Tr e^{(1-u)H}B e^{uH}A g_\alpha(u)\,du\,,
$$
where 
\begin{equation}
g_\alpha(u)=\frac{1}{\alpha(1-\alpha)}
\cases{ u  & if $0 \le u \le \alpha$,  
\cr  \alpha & if $\alpha \le u \le 1-\alpha$,
\cr 1-u & if $1-\alpha \le u \le 1$}
\end{equation}
for $\alpha \le 1/2$ and  for $\alpha \ge 1/2$ $g_\alpha=g_{1-\alpha}$.

Since
$$
\frac{\partial^2}{\partial t \partial s}S_\aa (e^{H+tA}|| e^{H+sB})=
{1 \over \aa(1-\aa)} \frac{\partial^2}{\partial t \partial s}
\Tr \exp \aa(H+sB) \exp (1-\aa)(H+tA)\,,
$$
we calculate as follows:
\begin{eqnarray*}
&&-{1 \over \aa(1-\aa)}\Tr \frac{\partial}{\partial s}\exp \aa(H+sB) 
\frac{\partial}{\partial t}\exp (1-\aa)(H+tA) \cr && \quad =
\Tr \int_0^1  \int_0^1 \exp(x\aa H) B \exp(1-x)\aa H
\exp(y(1-\aa) H) A \exp(1-y)(1-\aa) H\,dx dy
\cr && \quad =
\Tr \int_0^1  \int_0^1 \exp\Big((x\aa+(1-y)(1-\aa))H\Big) B 
\exp\Big(((1-x)\aa + y(1-\aa)) H\Big) A \,dx dy
\cr && \quad =
\Tr \int_0^1  \int_0^1 \exp\Big((x\aa- y+ y\aa-\aa+1)H\Big) B 
\exp\Big(-x\aa+ y- y\aa+\aa) H\Big) A \,dx dy
\cr && \quad =
\int_0^1  \int_0^1 F(-x\aa+ y- y\aa+\aa) \,dx dy
\end{eqnarray*}
for the functional 
$$
F(t)= e^{(1-t)H} B  e^{tH}  A\,.
$$
 We continue
\begin{eqnarray*}
&& \int_0^1  \int_0^1 F(-x\aa+ y- y\aa+\aa) \,dx dy
\cr && \qquad
= \int_0^1  \int_0^1 F(-x\aa+ y(1-\aa)+\aa) \,dx dy
\cr && \qquad
= \int_{x=0}^1 \frac{1}{1-\aa}\int_{z=0}^{1-\aa} 
F(-x\aa+ z +\aa) \,dz dx
\cr && \qquad
= \frac{1}{\aa}\int_{w=-\aa}^0 \frac{1}{1-\aa}\int_{z=0}^{1-\aa} 
F(z -w) \,dz dw
\cr && \qquad
= \int_0^1 F(u) g_\alpha(u) \,du\,,
\end{eqnarray*}
where $g_\alpha$ is as above. \qed

$$
\frac{\partial^2}{\partial t \partial s}S_\aa (e^{H+tA+sB}|| e^{H})=
{1 \over \aa(1-\aa)} \Tr \left(\frac{\partial^2}{\partial t \partial s}
\exp (1-\aa)(H+tA+sB) \right)\exp (\aa H)\,,
$$
We know that
$$
\frac{\partial^2}{\partial t \partial s} \exp (H+tA+sB)=
\Big|_{t=s=0}=\int_0^1 \int_0^s e^{(1-s)H}B e^{(s-u)H}Ae^{uH} \,duds\,,
$$
therefore
$$
\frac{\partial^2}{\partial t \partial s} \exp (1-\aa)(H+tA+sB)=
(1-\aa)^2 \int_0^1 \int_0^s  e^{(1-s)(1-\aa)H}B e^{(s-u)(1-\aa)H}A e^{u(1-\aa)H}\,duds\,,
$$
therefore we obtain
$$
 \int_0^1 \int_0^s \Tr e^{[1-(s-u)](1-\aa)H}B e^{(s-u)(1-\aa)H}A \,duds
=\int_0^1(1-x)\Tr e^{[1-x](1-\aa)H}B e^{x(1-\aa)H}A \,dx
$$

If $\alpha=0$, then we have the Kubo-Mori inner product. \qed
\end{pl}

\section{Von Neumann algebras}

Let $\iM$ be a von Neumann algebra. Assume that it is in standard form,
it acts on a Hilbert space $\iH$, $\iP\subset \iH$ is the positive cone 
and $J:\iH \to \iH$ is the modular conjugation. Let $\ffi$ and $\omega$ 
be normal states with representing vectors $\Phi$ and $\Omega$ in the 
positive cone. For the sake of simplicity, assume that $\ffi$ and 
$\omega$ are faithful. This means that $\Phi$ and $\Omega$ are cyclic 
and separating vectors. The closure of the unbounded operator 
$A\Phi \mapsto A^*\Omega$ has a polar decomposition $J\Delta(\omega/ 
\ffi)^{1/2}$ and $\Delta(\omega/\ffi)$ is called {\bf relative modular 
operator}. $A\Phi$ is in the domain of $\Delta(\omega/\ffi)^{1/2}$ 
for every $A \in \iM$.

For $A \in \iM$ and $f:\bbbr^+ \to \bbbr$, the quasi-entropy
\begin{equation}\label{E:quasi2}
S^A_f (\omega\| \ffi):= \< A \Phi , f(\Delta(\omega/\ffi))A\Phi \> 
\end{equation}
was introduced in \cite{PD26}, see also Chapter 7 in \cite{OP}. Of course,
(\ref{E:quasi}) is a particular case.
 
\begin{thm}\label{T:quasimon2}
Assume that $f:\bbbr^+ \to \bbbr$ is an operator monotone function
with $f(0)\ge 0$ and $\alpha:\iM_0 \to \iM$ is a Schwarz mapping. Then
\begin{equation}\label{E:quasimon2}
S^A_f (\omega\circ \alpha \| \ffi\circ \alpha) \ge
S^{\alpha(A)}_f (\omega \| \ffi)
\end{equation}
holds for $A \in \iM_0$ and for normal states $\omega$ and $\ffi$
of the von Neumann algebra $\iM$.
\end{thm}

The relative entropies are jointly convex in this setting similarly
to the finite dimensional case. Now we shall concentrate
on the generalized variance.

\subsection{Generalized covariance}

To deal with generalized covariance, we assume that 
$f:\bbbr^+ \to \bbbr$ is a standard operator monotone (increasing) function.
The natural extension of the covariance (from probability theory) is
\begin{equation}
\QCov^f_\omega(A,B)=\< \sqrt{f(\Delta(\omega/ \omega))}A \Omega, 
\sqrt{f(\Delta(\omega/ \omega))}B\Omega\>
-\overline{\omega(A)}\omega(B),
\end{equation}
where $\Delta(\omega/\omega)$ is actually the modular operator. Although
$\Delta(\omega/ \omega)$ is unbounded, the definition works. For the 
function $f$, the inequality
$$
\frac{2x}{x+1} \le f(x) \le \frac{1+x}{2}
$$ 
holds. Therefore $A\Omega$ is in the domain of $\sqrt{f(\Delta(\omega/
\omega))}$. 

For a standard function $f:\bbbr^+ \to \bbbr^+$ and for a normal
unital Schwarz mapping $\beta: \iN \to \iM$ the inequality
\begin{equation}
\QCov^f_\omega(\beta(X),\beta(X))
\le
\QCov^f_{\omega\circ \beta}(X, X) \qquad (X \in \iN)
\end{equation}
is a particular case of Theorem \ref{T:quasimon2} and it is the monotonicity 
of the generalized covariance under coarse-graining. The common symmetrized
covariance
$$
\Cov_\omega (A,B):= \fel \omega(A^*B+BA^*)-\overline{\omega(A)}\omega(B)
$$
is recovered by the particular case $f(t)=(1+t)/2$.

Since
$$
\QCov^f_\omega(A,B)=\gamma^f_\omega(A-\omega(A)I,B-\omega(B)I),
$$
it is enough to consider these sesquilinear forms on the
subspace $T_\omega:=\{ A \in\iM: \omega (A)=0\}$.

\subsection{The Cram\'er-Rao Inequality}

Let $\{\omega_\theta: \theta \in G\}$ be a smooth $m$-dimensional
manifold in the set of normal states of the von Neumann algebra $\iM$ and 
assume that a collection $A=(A_1,\dots,A_m)$ of self-adjoint operators is 
used to estimate the true value of $\theta$. The subspace spanned by
$A_1, A_2,\dots,A_m$ is denoted by $V$. 

Given a standard matrix monotone function $f$, we have the corresponding 
cost function
$$
\varphi_\theta [A,B]\equiv \QCov_f^{\omega_\theta} (A,B)
$$
for every $\theta$ and the cost  matrix of the estimator $A$ is a positive 
semidefinite matrix, defined by 
$$
\varphi_\theta [A]_{ij}=\varphi_\theta[A_i,A_j].
$$ 
The {\bf bias} of the estimator is
\begin{eqnarray*}
b(\theta)&=& \big(b_1(\theta), b_2(\theta),\dots, b_m(\theta)\big)\\
&:=&\big(\omega_\theta (A_1-\theta_1 I), \omega_\theta
(A_2-\theta_2 I), \dots, \omega_\theta (A_m-\theta_m I) \big).
\end{eqnarray*}
For an {\bf unbiased estimator} we have $b(\theta)=0$. From the
bias vector we form a bias matrix
$$
B_{ij}(\theta):= \partial_{\theta_i} b_j(\theta).
$$
For a {\bf locally unbiased estimator} at $\theta_0$, we have $B(\theta_0)=0$.

The relation
$$
\partial_{\theta_i} \omega_\theta (H) = \varphi_\theta [L_i(\theta),H] \qquad (H \in V)
$$
determines the {\bf logarithmic derivatives} $L_i(\theta)$. The {\bf Fisher
information matrix} is
$$
J_{ij}(\theta):= \varphi_\theta [L_i(\theta), L_j(\theta)].
$$

\begin{thm}\label{cr-rao} 
Let $A=(A_1,\dots,A_m)$ be an
estimator of $\theta$. Then for the above defined quantities
the inequality
$$
\varphi_\theta [A]\geq \big(I+B(\theta)\big) J(\theta)^{-1}
\big(I+B(\theta)^*\big)
$$
holds in the sense of the order on positive semidefinite matrices.
\end{thm}

Concerning the proof we refer to \cite{PD22}.

\subsection{Uncertainty relation}

In the von Neumann algebra setting the skew information (as a sesquilinear form) 
can be defined as
\begin{equation}\label{E:skew_vN}
I_\omega^f(X, Y):= 
\Cov_\omega(X,Y)-\QCov^{\tilde f}_\omega (X, Y)
\end{equation}
if $\omega(X)=\omega(Y)=0$. (Then $I_\omega^f(X)=I_\omega^f(X,X)$.)

\begin{lemma}\label{L:2}
Let $\iK$ be a Hilbert space with inner product $\bal \pont , \pont \jobb$ and
let $\< \pont ,\pont\>$ be a sesquilinear form on $\iK$ such that
$$
0 \le \< f,f\> \le \bal f, f \jobb
$$
for every vector $f \in \iK$. Then 
\begin{equation}\label{E:det}
[\, \< f_i,f_j\>\,]_{i,j=1}^m \le 
[\,\bal f_i, f_j \jobb \,]_{i,j=1}^m
\end{equation}
holds for every $f_1,f_2,\dots, f_m \in \iK$. 
\end{lemma}

\proof
Consider the Gram matrices $G:=[\,\bal f_i, f_j \jobb \,]_{i,j=1}^m$ and
$H:=[\, \< f_i,f_j\>\,]_{i,j=1}^m$, which are symmetric and
positive semidefinite. For every $a_1,\dots,a_m\in\bbbr$ we get
$$
\sum_{i,j=1}^m (\bal f_i, f_j \jobb - \< f_i,f_i\>) \overline a_ia_j
= \bal \sum_{i=1}^m a_if_i, \sum_{i=1}^m a_if_i \jobb
- \< \sum_{i=1}^m a_if_i, \sum_{i=1}^m a_if_i \> \ge 0
$$
by assumption. This says that $G-H$ is positive semidefinite, hence it is clear that
$G \ge H $. \qed

\begin{thm}\label{T:vegso}
Assume that $f,g:\bbbr^+\to \bbbr$ are standard functions 
and $\omega$ is a faithful normal state on a von Neumann algebra $\iM$.
Let $A_1,A_2, \dots ,A_m\in \iM$ be self-adjoint operators such that
$\omega(A_1)=\omega(A_2)= \dots =\omega(A_m)=0$. Then the determinant 
inequality
\begin{eqnarray}
&&\Det\biggl( \left[\QCov_{D}^g(A_i,A_j)\right]_{i,j=1}^m \biggr) 
\ge
\Det\biggl( \left[ 2g(0)I_\omega^f(A_i, A_j)\right]_{i,j=1}^m  \biggr)
\end{eqnarray}
holds.  
\end{thm}

\proof
Let $E(\pont)$ be the spectral measure of $\Delta(\omega, \omega)$. Then
for $m=1$ the inequality is
$$
\int g(\lambda) \,d\mu(\lambda)
\le g(0)\left(
\int \frac{1+\lambda}{2} \,d \mu(\lambda)
-\int {\tilde f}(\lambda) \,d \mu(\lambda)\right),
$$
where $d\mu(\lambda)=d \< A\Omega, E(\lambda)A\Omega\>$. Since the inequality
\begin{equation}\label{E:gibi}
f(x)g(x)\ge f(0)g(0) (x-1)^2
\end{equation}
holds for standard functions \cite{GHP}, we
have
$$
g(\lambda)\ge g(0)\left(\frac{1+\lambda}{2}-f(0){\tilde f}(\lambda)\right)
$$
and this implies the integral inequality.
 
Consider the finite dimensional subspace $\iN$ generated by the operators
$A_1,A_2, \dots ,A_m$. On $\iN$ we have the inner products
$$
\bal A, B \jobb :=\Cov_{\omega}^g(A,B)
$$
and
$$
\< A,B\>:= 2 g(0)I_\omega^f(A, B).
$$
Since $\< A,A\> \le \bal A, A \jobb$, the determinant inequality 
holds, see Lemma \ref{L:2}.\qed

This theorem is interpreted as quantum uncertainty principle
\cite{andai, GII, 3, Kosaki}. In the earlier works the function $g$ from the
left-hand-side was $(x+1)/2$ and the proofs were more complicated. 
The general $g$ appeared in \cite{GHP}.

\end{document}